\documentclass[prd,aps, floatfix,preprintnumbers]{revtex4}

\usepackage{tikz}
\usepackage{caption}
\usepackage{tikz}
\usepackage{graphicx}				
\usepackage{amsmath}
\usepackage{amssymb}
\usepackage{xspace}

\usepackage[applemac]{inputenc}
\usepackage[T1]{fontenc} 
 \usepackage{amsmath} 
\usepackage{amsfonts} 
\usepackage{amssymb, mathrsfs}

\def\beq{\begin{equation}}
\def\eeq{\end{equation}}
\def\bsp{\begin{split}}
\def\esp{\end{split}}
\def\bea{\begin{eqnarray}}
\def\eea{\end{eqnarray}}
\def\ba{\begin{array}}
\def\ea{\end{array}}

\def\l.{\left.}
\def\r.{\right.}

\begin{document}
\preprint{UdeM-GPP-TH-15-246}
\title{The phase-diagram of the Blume-Capel-Haldane-Ising spin chain}
\author{Christian Boudreault$^{a,c}$}
\email{Christian.Boudreault@cmrsj-rmcsj.ca}
\author{S. A. Owerre$^{a,b}$}
\email{alaowerre@gmail.com}
\author{M. B. Paranjape$^{a,b}$} 
\email{paranj@lps.umontreal.ca}
\affiliation{$^a$Groupe de physique des particules, D\'epartement de physique,
Universit\'e de Montr\'eal,
C.P. 6128, succursale centre-ville, Montr\'eal, 
Qu\'ebec, Canada, H3C 3J7 }
\affiliation{$^b$Perimeter Institute for Theoretical Physics, 31 Caroline Street North Waterloo, Ontario, Canada N2L 2Y5 }
\affiliation{$^c$Collge militaire royal de Saint-Jean,
Case Postale 100, succursale Bureau-chef,
Richelain, Qu\'ebec, Canada, J0J 1R0  }

\begin{abstract}

\section*{ABSTRACT}  
We consider the one-dimensional spin chain for arbitrary spin $s$ on a periodic chain with $N$ sites, the generalization of the chain that  was studied by Blume and Capel \cite{bc}: $$H=\sum_{i=1}^N \left(a (S^z_i)^2+ b S^z_iS^z_{i+1}\right).$$  The Hamiltonian only involves the $z$ component of the spin thus it is essentially an Ising \cite{Ising} model.  The Hamiltonian  also figures exactly as the anisotropic term in the famous model studied by Haldane \cite{haldane} of the large spin Heisenberg spin chain \cite{bethe}.   Therefore we call the model the Blume-Capel-Haldane-Ising model.  Although the Hamiltonian is trivially diagonal, it is actually not always obvious which eigenstate is the ground state.  In this paper we establish which state is the ground state for all regions of the parameter space and thus determine the phase diagram of the model.  We observe the existence of solitons-like excitations and we show that the size of the solitons depends only on the ratio $a/b$ and not on the number of sites $N$.  Therefore the size of the soliton is an intrinsic property of the soliton not determined by boundary conditions.
\end{abstract}



\maketitle

\section{Introduction}
Blume and Capel \cite{bc} studied a model that corresponded to an Ising spin chain with a nearest neighbour exchange interaction and an easy-plane/easy-axis interaction.  Ising spin chain means that only the $z$ component of the spin was dynamical,  and they only considered the model for spin 1.  The corresponding Hamiltonian was ab initio diagonal.  We will study the model for arbitrary large spin $s$.  The Hamiltonian is still diagonal.    The exchange interaction is minimized if neighbouring spins are maximal, $S_z=\pm s$,  and anti-aligned for positive coupling, but aligned for negative coupling.  The easy-plane/easy axis term, on the other hand,  is minimized if the spins are maximal, for negative coupling but minimal (zero if possible) for positive coupling.   Thus there can be a competition between the two terms and a variety of phases when one term or the other is dominant.  We are somewhat surprised to find second order phase transitions between the various phases described by the Hamiltonian.  Indeed, the two competing terms commute, and thus, in principle we expect only first order phase transitions.  However, here we find clearly that the critical points correspond to the appearance of negative eigenvalues of the energy functional, which is just a quadratic form, and the crossing of the ground state energy eigenvalue from positive to negative signals a second order phase transition.  

For antiferromagnetic coupling, with  odd $N$, there must be a defect in the Nel state on a periodic chain.  The defect is localized between two adjacent spins for easy-axis coupling, $a<0$.  However, for easy-plane coupling, $a>0$,  we find that the defect spreads out to maximal size as the anti-ferromagnetic coupling is weakened.  The defect is a soliton in the Nel state, and corresponds to a finite energy excitation.  We calculate the size of the soliton and find that it is independent of the number of sites $N$ but depends on the ratio $a/b$.  Hence the size is a characteristic that is not an artefact of the odd number of sites, the same solitons could be excited in the periodic chain with an even number of spins or the finite or infinite, open chain. 

Haldane \cite{haldane} considered the Heisenberg model with an anisotropy which corresponds exactly to the Blume-Capel model, with Hamiltonian
\beq
H=|J|\sum_{i=1}^N \vec S_i\cdot\vec S_{i+1} +\lambda S^z_iS^z_{i+1}+\mu (S^z_i)^2
\eeq
(with periodic boundary conditions, $\vec S_{N+1}=\vec S_1$), for large spin $|\vec S|=s\gg 1$ but for small anisotropy $0<(\lambda-\mu)^{1/2}\ll1$, $\lambda>\mu$.   However the complete phase diagram of the model, for all values of the couplings is still of much interest.  We define $a=|J|\mu$ and $b=|J|\lambda$, and consider the model for $|J|=0$ but $a$ and $b$ finite.  This is in the extreme opposite limit to Haldane's considerations, however, it is still of import to Haldane's considerations.  The anisotropy, however small in Haldane's work, picks the anti-ferromagnetic Nel ordered ground state that is aligned in the $z$ direction.  For different parts of the parameter space in the anisotropy, it is possible and indeed true that a different ground state is indicated.  

Thus we consider the Hamiltonian
\beq
H(S_1^z,\dots ,S_N^z)=\sum_{i=1}^{N} (a(S_{i}^z)^{2} + bS^z_{i}S^z_{i+1}).\label{h}
\eeq
This Hamiltonian is trivially  diagonal, the eigenstates can be written as $|s_1,\cdots ,s_N\rangle$, where $s_i\in\left[-s,-s+1,\cdots ,s-1,s\right]$ is the $z$ component of the $i$th spin, written in spin eigenstates that diagonalize the total spin and the $z$ component of the spin, for each individual spin.  The corresponding energy eigenvalue is $E(s_1,\cdots ,s_N)=\sum_{i=1}^{N} (a(s_{i})^{2} + bs_{i}s_{i+1})$.  Which eigenstate has the minimum energy  {\it i.e.} which is the ground state, is not always clearly evident.  
\section{Blume-Capel-Haldane-Ising model with arbitrary spin $s$}
\noindent 
We consider a one-dimensional periodic ``lattice'', a spin chain with $N$ sites $N\geq 2$ sites.  The Hamiltonian, \eqref{h} can be written as
\beq 
H=\frac{1}{2} \bold{S}^T A \bold{S}
\eeq 
where $\bold{S}^T=(S^z_1, S^z_2, \cdots, S^z_N)$.  

\subsection{The ferromagnetic case $(b<0)$ for all $N$.}
The ferromagnetic case is easily dealt with.  We write the state $|s_1,\dots ,s_N \rangle$ as $r |\alpha_1 , \dots , \alpha_N \rangle$ where $r=\sqrt{\sum_k s_k^2}$ and $\sum_k \alpha_k^2=1$.  Then the energy is given as 
\begin{equation}\label{E:spherical}
E(s_1,\dots ,s_N)=\big(a+b\textstyle\sum_k \alpha_k \alpha_{k+1}\big) r^2 = C(\boldsymbol{\hat{\alpha}})r^2
\end{equation}
where expicitly, 
\beq
C(\boldsymbol{\hat{\alpha}})=\big(a+b\textstyle\sum_k \alpha_k \alpha_{k+1}\big) .\label{C}
\eeq
As $\sum_k(\alpha_k\pm \alpha_{k+1})^2\ge 0$, we have 
\beq
-1\leq\sum_k \alpha_k \alpha_{k+1} \leq 1.\label{bound}
\eeq  
Thus for $|b| < a$, using Eqn.\eqref{bound} in Eqn.\eqref{C} we have $C(\boldsymbol{\hat{\alpha}})>0$ and hence the minimum energy configuration is realized exactly for $r=0$ corresponding to the state, $|0,\dots ,0\rangle$ with corresponding energy $E_0=0$.  For half odd integer spin, the state $|0,\dots ,0\rangle$ is not permitted. Then in this case, one of the states closest to the origin, $|\pm 1/2,\dots ,\pm 1/2\rangle$ (with uncorrelated $\pm$ signs) will be the minimal energy configuration.  Since the first term in the energy does not care whether the spin is $\pm 1/2$ and since $b<0$, the energy is minimized at the ``little'' ferromagnetic  states 
\beq
 |1/2,\dots ,1/2\rangle \quad {\rm or}\quad   |-1/2,\dots ,-1/2\rangle\label{littleferro}
 \eeq  
with energy $E_0=(1/4)N^2(a-|b|)$.

For  all other cases,  $|b|>a$, (including $a$ negative), the factor $C(\boldsymbol{\hat{\alpha}})$ becomes negative for certain directions, and in particular for $\boldsymbol{\hat{\alpha}}^T = \pm \frac{1}{\sqrt{N}} (1,\dots , 1)$  the upper bound Eqn.\eqref{bound}, for the sum $\sum_k \alpha_k \alpha_{k+1}$ is saturated.  These are the only two states for which the bound is saturated, and here $C(\boldsymbol{\hat{\alpha}})=a+b=a-|b|$.   $C(\boldsymbol{\hat{\alpha}})$ is negative and minimal for this direction.   The extreme corners  $\pm(s,\dots ,s)$ of the hypercube $[-s,s]^N$ are attained along this direction.  Hence $r$ is maximal, and correspondingly, the energy is minimal at the two corners.  Thus the two corresponding ferromagnetic states, which we will write as  $ | \uparrow, \dots ,\uparrow \rangle$ and $ | \downarrow, \dots ,\downarrow \rangle$, are the ground states in the regime $|b|>a$ (and we will use the notation $\uparrow$ and $\downarrow$ when the corresponding spin is maximally up, $s$, or maximally down, $-s$, respectively).

For $a=|b|$ the states   $\pm | m, \dots ,m \rangle$  are degenerate for any $m\in [-s,s]$ thus the ground state is $2s+1$ fold degenerate with $E_0=0$, and  the system passes through a highly degenerate critical point.  In the large $s$ limit this is veritably a massless continuum.

\subsection{The antiferromagnetic case $(b>0)$ for $N$ even}
The case of $N$ even and $N$ odd are substantially different, thus we will treat them separately.  For $N$ even the Nel state closes periodically as the lattice is bi-partite, but for $N$ odd, the Nel state does not close without frustration, one has a defect in the ground state. For $N$ odd we will see that the ground state is highly degenerate.

For $0<b<a$ continuing with the notation of Eqn. \eqref{E:spherical}, we find $C(\boldsymbol{\hat{\alpha}})>0$ for all directions $\boldsymbol{\hat{\alpha}}$.  Hence the minimum of the energy occurs at $r=0$, {\it i.e.} for the state $|0,\dots ,0\rangle$.  Therefore, for integer spin, the ground state is non-degenerate and given by $|0,\dots ,0\rangle$ with ground state energy $E_0=0$.  Now for half odd integer spins, this state is not permitted, and one of the closest nearby states, at the vertices of the hypercube of centred on the origin of side length 1, $|\pm 1/2,\dots ,\pm 1/2\rangle$ (the $\pm$ signs are not correlated) becomes the ground state.  For the case under consideration, $N$ even, it is easy to verify the minimum energy states correspond to the two ``little'' Nel states 
\beq
 |1/2,-1/2,1/2,-1/2,\dots ,1/2,-1/2\rangle \quad {\rm and}\quad  |-1/2,1/2,-1/2,\dots ,-1/2,1/2\rangle.\label{littleneel}
\eeq
Notice for this state, on the periodic lattice since $N$ is even, the Nel pattern closes without frustration.    Thus for half odd integer spin, the ground state is doubly degenerate and given by the two ``little'' Nel states, with ground state energy $E_0=(1/4)N^2(a-b)$.

For all other values, $a<b$ the factor $C(\boldsymbol{\hat{\alpha}})$ is negative in some directions, and in particular it is maximally negative for the states $\boldsymbol{\hat{\alpha}} = \pm \frac{1}{\sqrt{N}} (1,-1, \dots , 1, -1)$.   For these states the inequality Eqn.\eqref{bound}, which continues to hold for the anti-ferromagnetic case, is saturated with $\sum_k \alpha_k \alpha_{k+1}=-1$.  Then the energy is equal to $E=r^2 (a-b)$ which  is the ground state energy.  Thus the ground state of the system is doubly degenerate and corresponds to the two Nel states $ | \uparrow,\downarrow,\uparrow,\downarrow \dots ,\uparrow,\downarrow\rangle$ and $ | \downarrow,\uparrow,\downarrow,\uparrow, \dots ,\downarrow,\uparrow\rangle$.  
\subsubsection{Phase diagram for $N$ even }
\begin{figure}[ht]
\caption{Phase diagram for $N$ even and arbitrary spin $s$}\label{F:Npair}
\centering
\begin{tikzpicture}[scale=0.8, transform shape]
	\path [fill=black!5!white] (0.5,4.5) -- (9.5,4.5) -- (5,0) -- (0.5,0) -- (0.5,4.5);
	\path [fill=black!15!white] (5,0) -- (9.5,4.5) -- (9.5,-4.5) -- (5,0);
	\path [fill=black!25!white] (5,0) -- (9.5,-4.5) -- (0.5,-4.5) -- (0.5,0) -- (5,0);
	\draw [->] (0,0) -- (10,0);
	\draw [->] (5,-5) -- (5,5);
	\draw [dashed, ultra thick] (5,0) -- (9.5,4.5);
	\draw [dashed, ultra thick] (5,0) -- (9.5,-4.5);
	\draw (10,0) node[right] {$a$};
	\draw (5,5) node[right] {$b$};
	\draw (9.5,4.5) node[right] {$a=b=|b|$};
	\draw (9.5,-4.5) node[right] {$a=-b=|b|$};
	\path [fill=white] (7.8,2.7) rectangle (10,0.5);
	\draw (10.8,0.5) node[above] {$ \begin{matrix} 
					| 0, \dots , 0 \rangle 									& \text{integer spin}\\
					\left . \begin{matrix}
					|-\frac{1}{2}, \frac{1}{2} ,\dots , -\frac{1}{2},\frac{1}{2}\rangle \\
					 |\frac{1}{2}, -\frac{1}{2} ,\dots , \frac{1}{2},-\frac{1}{2}\rangle 	
					 \end{matrix} \right \}									& \text{half-odd spin}
						\end{matrix}$};
	\path [fill=white] (7.8,-2.7) rectangle (10,-0.6);
	\draw (10.6,-2.9) node[above] {$ \begin{matrix} 
					\left . \begin{matrix}
					|\frac{1}{2} ,\dots ,\frac{1}{2}\rangle \\
					 | -\frac{1}{2} ,\dots , -\frac{1}{2}\rangle 	
					 \end{matrix} \right \}									& \text{half-odd spin}\\
					| 0, \dots , 0 \rangle 									& \text{integer spin}\\
						\end{matrix}$};
	\path [fill=white] (2.5,1.5) rectangle (5.5,3);
	\draw (4,1.5) node[above] { $ \begin{matrix} 
							|\uparrow, \downarrow, \dots , \uparrow, \downarrow\rangle\\
							|\downarrow, \uparrow ,\dots , \downarrow ,\uparrow\rangle
						\end{matrix}$};
	\path [fill=white] (2.5,-3) rectangle (5.5,-1.5);
	\draw (4,-1.3) node[below] {$ \begin{matrix} 
							|\uparrow, \dots , \uparrow \rangle\\
							|\downarrow ,\dots , \downarrow\rangle
						\end{matrix}$};
\end{tikzpicture}
\end{figure}
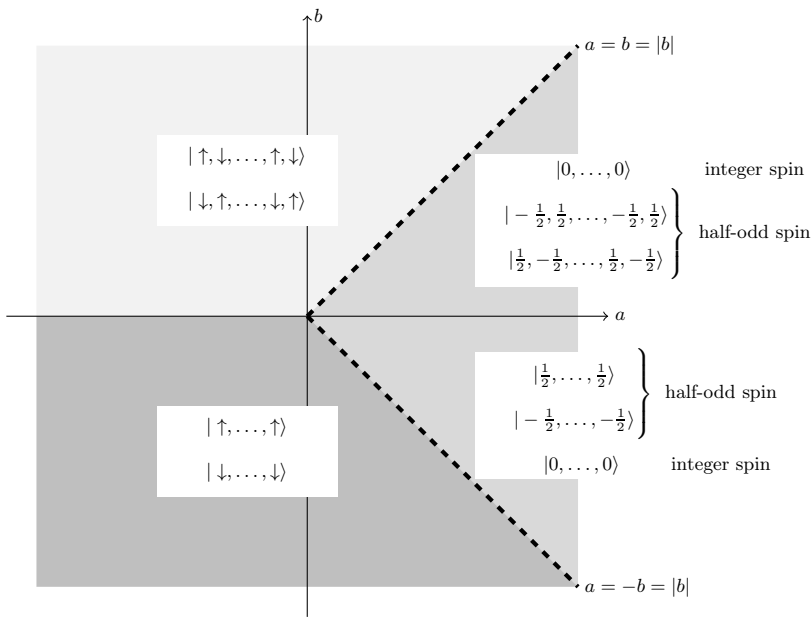

Fig. \eqref{F:Npair} gives the phase diagram for the case $N$ even for all $a$ and $b$.  For half odd integer spin the regions $a>|b|$ is  doubly degenerate with the ground states given by Eqn.\eqref{littleneel} or Eqn.\eqref{littleferro} depending on whether $b$ is positive or negative but non-degenerate with ground state $|0,0,\dots,0\rangle$ for integer spin.

We furthermore note that defining the staggered spin operators $\bar S_j\equiv (-1)^j S_j$ we find a mapping between the ferromagnetic and the anti-ferromagnetic cases
\beq 
	H^{b>0} (\bar{S}_1, \dots , \bar{S}_N) = H^{b<0} (S_1, \dots , S_N)\label{st}
\eeq 
and vice versa.  There is an obvious symmetry in the phase diagram between $b\to-b$.  The ground states in the two cases are related by the transformation $S_j \longleftrightarrow \bar{S}_j$, which can be seen from the Fig. \eqref{F:Npair}.

\subsection{The antiferromagnetic case $(b>0)$ for $N$ odd}
For the case of $N$ odd, the most interesting phenomena takes place for $a<b$, (note $b>0$ here).  For $N$ even this case had the ground states corresponding to the Nel states.  But now, the closest states to  Nel states correspond to  configurations with adjacent spins maximally anti-aligned, $(s,-s,s,-s,\cdots,s)$ or $(-s,s,-s,s,\cdots,-s)$ except for the last spin with the first spin since $N$ is odd.   For such a state, $\sum_k \alpha_k \alpha_{k+1} = -1 + \frac{2}{N} > -1$.  These are the ground states for $a<0$.  However, we will find that these states are not always the ground states.  We will show that  for $a>0$, as the ratio $a/b$ changes, the direction which minimizes $C(\boldsymbol{\hat{\alpha}})$ does not correspond to one of the corners of the hypercube and the minimum of the energy is attained for a value of $r$ that is not maximal.  As $a$ is increased from $a=0$, the ground state $r \boldsymbol{\hat{\alpha}}$ changes discretely to $r' \boldsymbol{\hat{\alpha}}'$ with $r'<r$.  The final transition occurs for $a\lessapprox b$ when the minimum value of $C(\boldsymbol{\hat{\alpha}})$ becomes positive and the ground state becomes $|0, \dots , 0\rangle$ for integer spins and the corresponding little Nel states $|1/2,-1/2,\dots,1/2,-1/2,1/2\rangle$ and $|-1/2,1/2,\dots,-1/2,1/2,-1/2\rangle$, for half odd integer spin.

The energy of a general state $|s_1,\dots ,s_N\rangle$ is of course given by
\beq 
E(s_1,\dots ,s_N)=\sum_{i=1}^{N} (as_{i}^{2} + bs_{i}s_{i+1})=\frac{1}{2} \bold{s}^T A \bold{s}.
\eeq 
The critical points are obtained by solving the linear system of equations
\beq
\frac{\partial E(s_1,\dots ,s_N)}{\partial s_i}=(2as_{i} + b(s_{i+1}+s_{i-1})=\sum_{j=1}^NA_{ij}s_j=0,
\eeq
which has the evident solution $s_i=0$ for all generic points where $A$ is invertible.  Then the corresponding state $|0,\dots ,0\rangle$ will be the ground state if the energy is at a minimum at this point.  The  matrix $A$, which is the Hessian matrix of second derivatives of the energy with respect to $s_i$ and $s_j$, is a circulant matrix, \cite{dav}, with first row $(2a,b,0,\dots , 0,b)$.  Subsequent rows correspond to the previous row permuted circularly by one, so for example, the second row is $(b,2a,b,0,\dots , 0)$, and so on.  The eigenvalues and eigenvectors of circulant matrices are well known, the eigenvalues are $\lambda_k=2(a+b\cos\frac{2\pi k}{N})$, $k=0,1,\dots ,N-1$ with corresponding normalized eigenvectors $\bold s_k^T=\frac{1}{\sqrt N}(1,\omega_k,\omega_k^2,\dots,\omega_k^{N-1})$ where $\omega_k$ is the $k$th, $N$th root of unity, $\omega_k=e^{\frac{i2\pi k}{N}}$.  The eigenvalues are doubly degenerate except for $\lambda_0$.  For $N$ odd, the Hessian is a positive definite matrix for the entire region $a>b\cos\frac{\pi}{N}$, which corresponds to the point at which the smallest eigenvalue at $k=[N/2]$ or $k=[N/2]+1$  vanishes and then becomes negative for smaller $a$.  (For $N$ even,  $k=N/2$ is allowed and the Hessian is positive definite only for $a>b$.) Thus for $a>b\cos\frac{\pi}{N}$, the minimum of the energy is trivially obtained to be at $|0,\dots ,0\rangle$, which is then, of course, the ground state for integer spin.  For half odd integer spin, the ground state must be chosen from the corners of the (smallest) hypercube of side length 1 centered on the origin, with vertices $(\pm1/2,\pm1/2,\dots\pm1/2)$, with uncorrelated $\pm$ signs.  It is easy to see that then the little Nel states, with one defect, chosen for example between the first and last sites, $|1/2,-1/2,\dots,1/2,-1/2,1/2\rangle$ and $|-1/2,1/2,\dots,-1/2,1/2,-1/2\rangle$, correspond to the ground states.  Each one is then in fact $N$ fold degenerate, since the position of the defect can be placed at any of the $N$ different sites.

For the region $0<\frac{a}{b}< \cos\frac{\pi}{N}$, the Hessian is indefinite, and the critical point at $s_i=0$ is a saddle point.  Thus the minimum energy is not attained at the critical point.  Then Fermat's theorem for the extrema of a differentiable function defined on a compact set implies that, since the minimum does not occur and an internal point,  it must occur on the boundary of the hypercube and we must look for the minimum on its surface.  Without loss of generality, we go to the boundary by taking  $s_N = s$.  Taking $s_N = -s$ is clearly also possible, but corresponds simply to the solution that we will find if we flip all the spins, $s_i\to -s_i$. The  energy, taking the face  $s_N = s$, is then
\beq
E=a\sum_{i=1}^{N-1} s_i^2 +as^2+b\sum_{i=1}^{N-2}s_is_{i+1}+bs(s_{N-1}+s_1).
\eeq
Varying with respect to $s_i$ to find the critical points gives the set of equations
\bea
2a s_i+b(s_{i+1}+s_{i-1})=0\quad i=2,\cdots ,N-2,\cr
2a s_1+b(s_2+s)=0 ,\quad{\rm and}\quad 2a s_{N-1}+b(s_{N-2}+s)=0.\label{crit}
\eea
These can be easily solved, however it is not particularly useful, since the corresponding Hessian is not positive definite.  The Hessian is an $N-1\times N-1$ matrix, which we will call  $B_{N-1}$, which is a tridiagonal Toeplitz matrix with the three non-zero diagonals given by
\beq\label{B}
	B_{N-1}=\left(\begin{matrix}
					 \vspace{-.3cm}

		 2a& b & &  &  &  &&  \\
		 				 \vspace{-.3cm}

		b&2a  & b &  &  &  & &\\
				 \vspace{-.3cm}

		 & b &2a  & b &  &  & & \\
		 \vspace{-.3cm}
		 &&\cdot&\cdot&\cdot&&&\\
		 		 \vspace{-.3cm}
		 &&&\cdot&\cdot&\cdot&&\\
		 				 \vspace{-.3cm}

		 &&&&\cdot&\cdot&\cdot&\\
		 				 \vspace{-.3cm}
&&&&&b&2a&b\\
		 &&&&&&b&2a\\
	\end{matrix}
	\right)_{(N-1)\times (N-1)}
\eeq
Again it is well understood how to find the eigenvalues and eigenvectors of a Toeplitz matrix, \cite{toeplitz}.  The eigenvalues are given by $\lambda_k = 2\big(a+b\cos \frac{k\pi}{N}\big)$, $k=1,2, \dots , N-1$.  Thus, this matrix is  again indefinite in the entire region $0<\frac{a}{b}< \cos\frac{\pi}{N}$ and we must go to the boundary of the face, $s_N=s$, of the hypercube by setting one of the remaining $s_i$'s equal to $s$.  This $s_i$ can be chosen adjacent to $s_N$ or separated from it.  We consider first the case $s_N=s$, $s_{N-1}=\pm s$ and return to the separated case later.  

Taking $s_N=s$, $s_{N-1}= -s$ we get the energy
\beq
E=a\sum_{i=1}^{N-2} s_i^2 +2as^2+b\sum_{i=1}^{N-3}s_is_{i+1}+bs(-s+s_1)-bss_{N-2}.
\eeq
with the corresponding critical point given by the, now inhomogeneous equations, analogous to Eqn. \eqref{crit},
\beq
B_{N-2}\bold s=-b\bold t\label{crit1}
\eeq 
where $B_{N-2}$, which is as defined in Eqn.\eqref{B}, is the Toeplitz matrix of dimension $N-2\times N-2$, as usual $\bold s = (s_1, s_2, \dots , s_{N-2})^T$ and $\bold t = (s,0,\dots ,0,-s)^T$.  The equation for the critical point can be generally solved however, this is again not necessary for the full region $0<\frac{a}{b}< \cos\frac{\pi}{N}$, since the Hessian matrix is $B_{N-2}$ whose eigenvalues and eigenvectors are easily found.  The Hessian is $B_{N-2}$  with eigenvalues $\lambda_k = 2\big(a+b\cos \frac{k\pi}{N-1}\big)$, $k=1,2, \dots , N-2$.  Now the Hessian is positive definite for $\cos\frac{\pi}{N-1}<\frac{a}{b}<\cos\frac{\pi}{N}$, therefore in this range the solution for the critical point, Eqn.\eqref{crit1} is the minimum energy configuration.  The solution is obtained by inverting the Toeplitz matrix $B_{N-2}$, $\bold s=-b(B_{N-2})^{-1}\bold t$, explicitly from \cite{toeplitz}
\begin{equation}\label{sk}
	s_k = (-1)^k s \bigg( \frac{\sin (N-1-k)\theta - \sin k\theta}{\sin (N-1)\theta}\bigg),\qquad\qquad k=1,\dots , N-2,
\end{equation}
where $\cos \theta = a/b$.   For $\frac{\pi}{N}<\theta <\frac{\pi}{N-1}$

\beq 
	\begin{aligned}
	\bigg|\frac{\sin (N-1-k)\theta - \sin k\theta}{\sin (N-1)\theta}\bigg| &=\frac{|2\sin \frac{(N-1-2k)\theta}{2} \cos\frac{(N-1)\theta}{2}|}{2\sin \frac {(N-1)\theta}{2}\cos \frac {(N-1)\theta}{2}}\\
													&=\frac{|\sin \frac{(N-1-2k)\theta}{2} |}{\sin \frac {(N-1)\theta}{2}}=\frac{|\sin \left(\frac{N-1}{2}-k\right) \theta|}{\sin \left(\frac {N-1}{2}\right)\theta}\\
													&<\frac{|\sin \left(\frac{N-1}{2}-1\right) \frac{\pi}{N-1}|}{\sin \left(\frac {N-1}{2}\right) \frac{\pi}{N}}=\frac{|\sin \left(\frac{\pi}{2}-\frac{\pi}{N-1}\right) |}{\sin \left(\frac {\pi}{2}-\frac{\pi}{2N}\right) }\\
													&=\frac{|\cos \left(\frac{\pi}{N-1}\right) |}{\cos \left(\frac{\pi}{2N}\right) }\\
													&<1,
	\end{aligned}
\eeq 
where in the first equality we have used a standard trigonometric identity, then we have used the fact that the expression is symmetric for $k\to N-1-k$, hence we can restrict $k\in1,2,\dots,\left[\frac{N-1}{2}\right]$ on which domain the sine function is monotone increasing.  Thus at the first inequality we replace the argument in the numerator with the largest possible value and the argument in the denominator with its smallest value and the second inequality is obvious.  Therefore we have $| s_k | < s$. 
 
For the case $s_N=s=s_{N-1}$ (or with $s_{N-1}=-s$ for $N$ \textit{even}), a similar calculation gives
\beq 
	s_k = (-1)^k s \bigg( \frac{\sin (N-1-k)\theta + \sin k\theta}{\sin (N-1)\theta}\bigg),\qquad\qquad k=1,\dots , N-2.
\eeq 
Now the sum in the numerator, instead of the difference that we had before, forces $| s_k |>s$, which is not permitted, therefore this solution is unacceptable.  However it is easy to determine the lowest energy configuration subject to the boundary condition $s_N=s=s_{N-1}$.  This boundary condition makes the chain completely equivalent to an open chain obtained by cutting the periodic chain between the $N$th and $N-1$th sites, and imposing the boundary condition that the spins at the ends are equal to $s$.  The contribution to the energy of these end spins is just a constant hence the minimum energy configuration of this chain will not be affected.  We can easily prove using mathematical  induction that, for an open chain of length $N$, even or odd, the Nel states are the states of minimum energy for the parameter range that we are in, $\frac{\pi}{N}<\theta <\frac{\pi}{N-1}$, when no boundary condition is imposed.  Thus for an odd number of sites, there is a Nel state that will satisfy the boundary condition $s_N=s=s_{N-1}$.  This chain has an odd total number of spins and subject to these boundary conditions the Nel configuration fits exactly without defect and the state  $|\downarrow\uparrow \dots \downarrow\uparrow\uparrow\rangle$  corresponds to the configuration of minimum energy subject to the boundary condition.  We note of course that this state is not the ground state of the periodic chain with the same number of spins, as one can easily check, for example,  $|\downarrow\uparrow \dots \downarrow\uparrow 0\rangle$ has lower energy (as $a>0$).  The actual ground state corresponds to the boundary conditions $s_N=s$,  $s_{N-1}=-s$ as analyzed first, with the soliton of Eqn.\eqref{sk} interpolating through the spins 1 to $N-2$,   and of course the states obtained by translation of the soliton.

It will be useful in our subsequent analysis to record here what happens for the case of even number of spins.  Here, with  boundary condition $s_N=s$ et $s_{N-1}=-s$, it follows that the minimum energy configuration is the corresponding Nel state, it fits without frustration.  However with the boundary condition $s_N=s=s_{N-1}$ for $N$ even, it is an easy repetition of the previous analysis to see that we get the same solution $s_k$ as in Eqn. \eqref{sk}, but now since $N$ is even, the first and last spins both point in the same direction.   We summarize our finding in Table \eqref{T:contraintes}.

\begin{table}[h]
\caption{Ground states for the open chain with boundary conditions.}\label{T:contraintes}	
	\begin{center}
		\begin{tabular}{ | r | c | c | }
		\hline
					& $|\uparrow, s_2,\dots , s_{N-1}, \downarrow\rangle$		& $|\uparrow, s_2,\dots , s_{N-1}, \uparrow\rangle$	\\ \hline
		$N$ odd	& soliton Eqn.(\ref{sk})								& Nel	\\ \hline
		$N$ even		& Nel											&  soliton Eqn.(\ref{sk})	\\ \hline
		\end{tabular}
	\end{center}
\end{table}
Thus the case where the second spin is adjacent to the first one is completely understood. If the two spins are up, for a chain with an odd number of sites, the rest of the chain is in the appropriate Nel state, while if the chain has an even number of sites, the rest of the chain is in the soliton configuration given by of Eqn.\eqref{sk}.  If the two spins are opposite then for an odd number of sites we find the soliton state of Eqn.\eqref{sk} and for an even number of sites we have the appropriate Nel state.

Next we will consider, and eliminate, the case where the second spin is not adjacent to the first fixed spin.  We first found that we must go to the boundary, the face corresponding to taking $s_N=s$, but then we found we must go to the boundary of this face.  We will now do so by taking $s_j=s$ with specifically $j\ne N-1$ or $1$, so that the second fixed spin is not adjacent to the first.  Thus we fix $j$ to be an integer in the interval $[2,N-2]$ and put ${s'}_N=s$ and ${s'}_{N-j}= -s$ (the case ${s'}_{N-j}= +s$ will follow).  This will give rise to a $(N-2)\times (N-2)$ system of equations, the solution of which will be the appropriate critical point if it corresponds to a minimum.  Thus we will first consider the Hessian at this point, which is a matrix that is block diagonal
\beq 
	C_{N-2}=\left(
	\begin{matrix}
		 B_{N-j-1} & 0 \\
		 0 		& B_{j-1} \\
	\end{matrix}
	\right)=B_{N-j-1} \oplus B_{j-1}\: ,
\eeq 
the eigenvalues are given by $\lambda_k = 2\big(a+b\cos \frac{k\pi}{N-j}\big)$, $k=1,2, \dots , N-j-1$ and $\lambda_k' = 2\big(a+b\cos \frac{k'\pi}{j}\big)$, $k'=1,2, \dots , j-1$.  It is a straightforward exercise to verify that these are  positive definite as we are in the region  $\cos\frac{\pi}{N-1}<\frac{a}{b}<\cos\frac{\pi}{N}$.  Thus the critical point is the minimum and designates the putative ground state, which we identify through the following analysis.  The spin ${s'}_{N-j}=-s$ divides the sequence $|{s'}_1 , \dots , {s'}_{N-j-1},-s,{s'}_{N-j+1},\dots ,s\rangle$ into two parts, one of which must be even.  Without loss of generality, we take $|-s, {s'}_{N-j+1},\dots ,s\rangle$ sequence to be even.  Then. consulting Table \eqref{T:contraintes}, the energy is minimized on this segment by the Nel state, $|-s,+s,\dots , -s,+s\rangle$, and the minimum energy sequence on the remaining odd length segment is given by the solution $\bold{s'}=({s'}_1,\dots , {s'}_{N-j-1})^T$ of the equation
\beq 
	B_{N-j-1} \: \bold{s'} = -b\bold{t},
\eeq 
where $\bold{t}=(s,0,\dots ,0,-s)^T$ with solution as before given by
\beq 
	{s'}_k = (-1)^k s \bigg( \frac{\sin (N-j-k)\theta - \sin k\theta}{\sin (N-j)\theta}\bigg), \qquad k=1,\dots ,N-j-1.
\eeq 
But this state, $|{s'}_1 , \dots ,{s'}_{N-j-1},-s,+s,\dots , -s,+s\rangle$ cannot be the lowest energy state. Since $s'_{N-1}=-s$, and we have already found the state that minimizes the energy when $s_N=s=-s_{N-1}$, the true minimum energy state $|\bold{s}\rangle$ is given by   Eqn. \eqref{sk}.   Thus the case where the second fixed spin $s_{N-j}=-s$, $j\in [2,N-2]$ is not adjacent to $s_N=s$ gives a higher,  minimum energy state than that obtained when the second spin is adjacent, $j=1$.  Therefore the second spin must be adjacent to the first.  

The case $s_{N-j}=+s$, $j\in [2,N-2]$ is eliminated by an essentially identical analysis.  We can now take the segment  $|s, {s'}_{N-j+1},\dots ,s\rangle$ to be odd. Then we find the state $|{s'}_1 , \dots ,{s'}_{N-j-1},+s,-s,\dots , -s,+s\rangle$ minimises the energy with this constraint. Again, this state satisfies $s_{N-1}=-s$ et $s_N=s$ and it has already been found that it is not the minimal energy configuration.  Thus we have eliminated  the possibility that the second fixed spin is not adjacent to the first fixed spin, and we have shown that the ground state is simply given by the solution in Eqn. \eqref{sk}.

Now we can easily see what happens as  more levels become negative.  In the region $\cos\frac{\pi}{N-2}<\frac{a}{b}<\cos\frac{\pi}{N-1}$ another eigenvalue becomes negative, and we must again go to the boundary to look for the minimum energy configuration.  If we take $s_N=s, s_{N-1}=-s$ and $s_{N-2}=s$ the Hessian $B_{N-3}$ will be positive definite and the minimum will be obtained as before as for the even chain with boundary values $+s$ on each end.  It is also clear that as $\frac{a}{b}$ is further reduced, each time  its value passes $\cos\frac{\pi}{m}$, for  the integers $m=N, N-1, \dots , 3$ will cause one more negative energy mode, and then require the minimum to be found by again ``going to the boundary''.  Each time, a new spin with $s_{N-m}=\pm s$ will attach to the sequence of spins.    For $m=N, N-1, \dots , 4$, the absolute minimum in the region  $\cos\frac{\pi}{m-1}<\frac{a}{b}<\cos\frac{\pi}{m}$ will be obtained as before by solving the system with Hessian $B_{m-2}$. 

\begin{figure}[h]
\caption{Ground state for $N=7$ as $a/b$ is varied.}\label{picture2}
\begin{tikzpicture}
	\draw [->] (0,0) -- (13.5,0);
	\draw (0,0) node[below] {$0$} node[above] {$|$};
	\draw (2,1) node[above,rotate=30] {$|0,\downarrow,\uparrow,\downarrow,\uparrow,\downarrow,\uparrow\rangle$};
	\draw (2.5,0) node[below] {$b\cos\frac{\pi}{3}$} node[above] {$|$};
	\draw (4.5,1) node[above,rotate=30] {$|s^2_1,s^2_2,\uparrow,\downarrow ,\uparrow,\downarrow,\uparrow\rangle$};
	\draw (5,0) node[below] {$b\cos\frac{\pi}{4}$} node[above] {$|$};
	\draw (7,1) node[above,rotate=30] {$|s^3_1,s^3_2,s^3_3,\downarrow,\uparrow,\downarrow,\uparrow\rangle$};
	\draw (7.5,0) node[below] {$b\cos\frac{\pi}{5}$} node[above] {$|$};
	\draw (9.5,1) node[above,rotate=30] {$|s^4_1,s^4_2,s^4_3,s^4_4,\uparrow,\downarrow,\uparrow\rangle$};
	\draw (10,0) node[below] {$b\cos\frac{\pi}{6}$} node[above] {$|$};
	\draw (12,1) node[above,rotate=30] {$|s^5_1,s^5_2,s^5_3,s^5_4,s^5_5, \downarrow,\uparrow\rangle$};
	\draw (12.5,0) node[below] {$b\cos\frac{\pi}{7}$} node[above] {$|$};
	\draw (14,0.5) node[above,rotate=30] {$|0,\dots ,0\rangle$};
	\draw (13.5,0) node[right] {$a$};
\end{tikzpicture}
\end{figure}
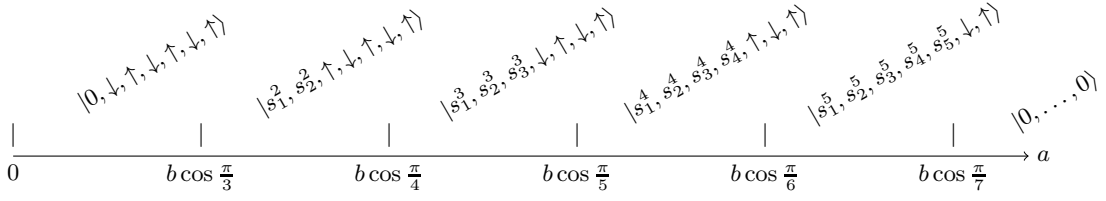

 For the final domain, $0<\frac{a}{b}<\cos\frac{\pi}{3}$, the problem is trivial and gives the ground state $|0,-s,+s,\dots , -s,+s\rangle$.  We confirmed these results numerically for  $N=5,7$ et 9. The picture of the ground state which emerges  is shown for $N=5$ in Fig. \eqref{picture1}, where $\uparrow$ and $\downarrow$ represent $s$ et $-s$, respectively. The similar picture for $N=7$ is given in Fig. \eqref{picture2}.
 
 It is important to note that the Hessian $B_{m-2}$ and the solution $s^{m-2}_k$ are independent of $N$, and depend only on $a/b$ and $m$.  For fixed $a/b$, the ground state of systems of different sizes only differ by the length of the part of the chain that is in the Nel configuration.   Thus in particular, the size of the soliton depends only on $a/b$ and is independent of $N$.  Hence the soliton is universal and will appear unchanged as an excitation in any spin chain, be it one with an even number of sites, or an open chain, finite or infinite.

\begin{figure}[h]
\caption{Ground state for $N=5$ as $a/b$ is varied.}\label{picture1}
\begin{tikzpicture}
	\draw [->] (0,0) -- (12,0);
	\draw (0,0) node[below] {$0$} node[above] {$|$};
	\draw (1.5,0.5) node[above] {$|0,\downarrow,\uparrow,\downarrow,\uparrow\rangle$};
	\draw (3,0) node[below] {$b\cos\frac{\pi}{3}$} node[above] {$|$};
	\draw (4.5,0.5) node[above] {$|s^2_1,s^2_2,\uparrow,\downarrow ,\uparrow\rangle$};
	\draw (6,0) node[below] {$b\cos\frac{\pi}{4}$} node[above] {$|$};
	\draw (7.5,0.5) node[above] {$|s^3_1,s^3_2,s^3_3,\downarrow,\uparrow\rangle$};
	\draw (9,0) node[below] {$b\cos\frac{\pi}{5}$} node[above] {$|$};
	\draw (10,0.5) node[above] {$|0,\dots ,0\rangle$};
	\draw (12,0) node[right] {$a$};
\end{tikzpicture}
\end{figure}
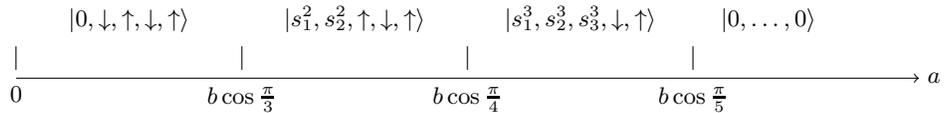

It remains to analyze the case $a<0$. We will consider to the energy function  $E(s_1,\dots ,s_N)=\big(a+b\sum_k \alpha_k \alpha_{k+1}\big) r^2 = C(\boldsymbol{\hat{\alpha}})r^2$and the change in the ground state as the parameter $a$ decreases from  $a>b$ to $a<0$. As $a$ first drops below $b\cos\pi/N$, which of course is very close to $b$ if $N$ is large, the factor $C$ becomes negative in the direction $\boldsymbol{\hat{\alpha}}$ corresponding to the solitons of length $N-2$. The energy is minimized at the intersection of the direction $r\boldsymbol{\hat{\alpha}}$ and the boundary of the cube $[-s,s]^N$. This intersection corresponds exactly to the soliton of Eqn.(\ref{sk}) and the  Nel part contains just two spins. As $a$ passes next below $b\cos\pi/(N-1)$, the system gets a new ground state  $| r' \boldsymbol{\hat{\alpha}}' \rangle$, which is the soliton of length $N-3$ with a Nel part containing three spins.  For the new ground state, $r'>r$ and $\frac{\partial E'}{\partial a}=r'^2>\frac{\partial E}{\partial a}=r^2$. The same thing happens each time $a$ passes $b\cos\pi/m$, and each time $r$ increases until it reaches $s\sqrt{N-1}$ in the direction $\pm \frac{1}{\sqrt{N-1}}(1,-1,\dots , -1,0)$ when $a\gtrapprox 0$ (see Fig. \eqref{F:Nimpair}) explicitly $0<a<b\cos\frac{\pi}{3}$.    

In the end, we show that
$|\mbox{Nel} + \mbox{1 defect}\rangle$ becomes the ground state for $a<0$. This is obviously the case for $a\to -\infty$.  For finite values, it suffices to calculate the energy values directly  for
$| \uparrow,\downarrow,\dots ,\uparrow,\downarrow, 0\rangle$ and $|\mbox{Nel} + \mbox{1 defect}\rangle$, we get 
\beq 
E = (N-1)as^2 -(N-2)bs^2 \qquad\mbox{and}\qquad E' = Nas^2 - (N-2)bs^2,
\eeq 
respectively. We see the two will cross exactly at $a=0$. Since the states $|\mbox{Nel} + \mbox{1 defect}\rangle$ saturates the  upper bound for $r$,  $s\sqrt{N}$, the derivative $\frac{\partial E}{\partial a}=s^2N$ is maximal. Thus there can be no intermediate transition of the type $|\uparrow,\downarrow,\dots, \uparrow,\downarrow,0\rangle \to | r\boldsymbol{\hat{\alpha}}\rangle \to |\mbox{Nel} + \mbox{1 defect}\rangle$.  It is also easy to prove this rigorously using mathematical induction.

\subsubsection{The phase diagram for $N$ odd.}
\begin{figure}[h]
\caption{Phase diagram for $N$ odd.}\label{F:Nimpair}
\centering
\begin{tikzpicture}[scale=0.8, transform shape]
	\path [fill=black!5!white] (0.5,4.5) -- (5,4.5) -- (5,0) -- (0.5,0) -- (0.5,4.5);
	\path [fill=black!15!white] (5,0) -- (9.5,4.5) -- (9.5,-4.5) -- (5,0);
	\path [fill=black!25!white] (5,0) -- (9.5,-4.5) -- (0.5,-4.5) -- (0.5,0) -- (5,0);
	\path [fill=black!55!white] (5,0) -- (9.1,4.5) -- (9.5,4.5) -- (5,0);
	\draw [->] (0,0) -- (10,0);
	\draw [->] (5,-5) -- (5,5);
	\draw [dashed] (5,0) -- (7.2,4.5); 
	\draw (8,5.5) node [above,rotate=64] {$a=b\cos\pi/3$};
	\draw [dashed] (5,0) -- (8.2,4.5); 
	\draw (9.2,5.5) node [above,rotate=56] {$a=b\cos\pi/4$};
	\draw [dashed] (5,0) -- (8.7,4.5);
	\draw (9.3,5) node [above] {$\dots$};
	\draw [dashed] (5,0) -- (8.9,4.5);
	\draw [dashed] (5,0) -- (9.1,4.5);
	\draw [dashed, ultra thick] (5,0) -- (9.5,4.5);
	\draw [dashed, ultra thick] (5,0) -- (9.5,-4.5);
	\draw (10,0) node[right] {$a$};
	\draw (5,5) node[right] {$b$};
	\draw (9.5,4.5) node[right,rotate=45] {$\boldsymbol{a=b\cos\pi/N}$};
	\draw (9.5,-4.5) node[right] {$a=|b|$};
	\path [fill=white] (7.8,2.7) rectangle (10,0.5);
	\draw (10.8,0.5) node[above] {$ \begin{matrix} 
					| 0, \dots , 0 \rangle 									& \text{integer spin}\\
					\left . \begin{matrix}
					|-\frac{1}{2}, \frac{1}{2} ,\dots , -\frac{1}{2},\frac{1}{2},-\frac{1}{2}\rangle \\
					 |\frac{1}{2}, -\frac{1}{2} ,\dots , \frac{1}{2},-\frac{1}{2},\frac{1}{2}\rangle 	
					 \end{matrix} \right \}									& \text{half-odd spin}
						\end{matrix}$};
	\path [fill=white] (7.8,-2.7) rectangle (10,-0.6);
	\draw (10.6,-2.9) node[above] {$ \begin{matrix} 
					\left . \begin{matrix}
					|\frac{1}{2} ,\dots ,\frac{1}{2}\rangle \\
					 | -\frac{1}{2} ,\dots , -\frac{1}{2}\rangle 	
					 \end{matrix} \right \}									& \text{half-odd spin}\\
					| 0, \dots , 0 \rangle 									& \text{integer spin}\\
						\end{matrix}$};
\path [fill=white] (1,1.5) rectangle (4.5,2.5);
	\draw (2.7,1.7) node[above] {$\pm |\mbox{Nel} + \mbox{1 defect}\rangle$};
	\path [fill=white] (2.5,-2.5) rectangle (5.5,-1.5);
	\draw (4,-1.45) node[below] {$ |\uparrow,\dots ,\uparrow\rangle$};
	\draw (4,-1.95) node[below] {$ |\downarrow,\dots ,\downarrow\rangle$};
	\draw [fill=white] (5.5,3) rectangle (7.8,3.8);
	\draw (6.7,3.1) node[above] {$|\mbox{soliton}\rangle$};
\end{tikzpicture}
\end{figure}
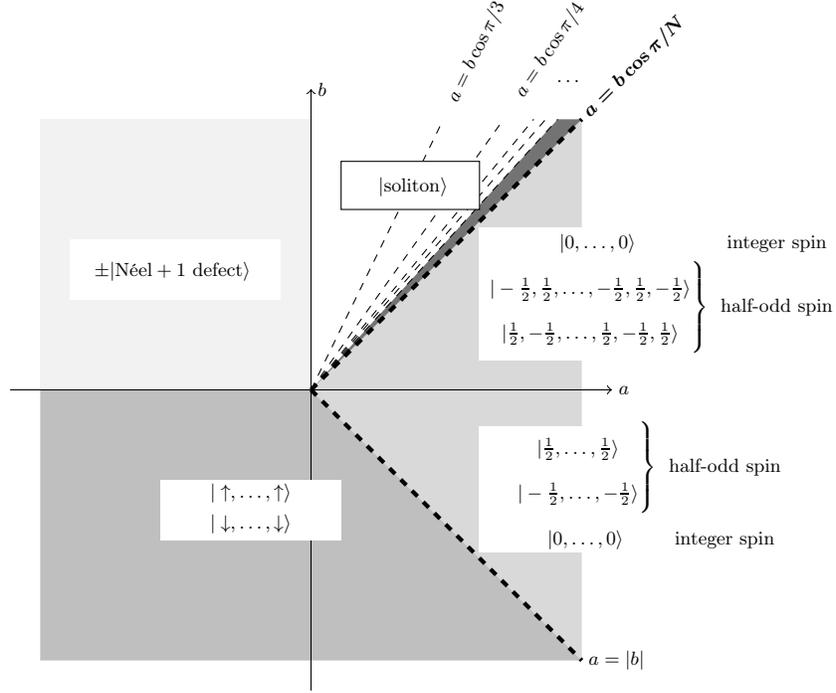

The phase diagram for $N$ odd is then given in Fig.\eqref{F:Nimpair}.  We  note that the duality that exists for $N$ even between the ferromagnetic and the antiferromagnetic cases is only partially valid for $N$ odd: 
\beq 
	H^{b>0} (\bar{S}_1, \dots , \bar{S}_N) = H^{b<0} (S_1, \dots , S_N)+2bS_N S_1
\eeq 
and vice versa. 
\vskip4cm
\section{Conclusion}  
We have found  the phase diagram for the Blume-Capel-Haldane-Ising model for a periodic chain with $N$ sites.  The Hamiltonian contains an easy-plane/easy-axis interaction and an exchange term,  with coefficients $a$ and $b$ respectively.  Although the Hamiltonian is trivial and the energy eigenstates are evident, it is not clear which eigenstate is the ground state.  In the ferromagnetic case, $b<0$, there is a straightforward competition between the exchange interaction and the easy-plane/easy axis interaction.  The ground state switches from ferromagnetic when all the spins are $s$ or all the spins are $-s$, to easy plane configuration, where all the spins are $s=0$, as we cross the phase boundary at $a=|b|$.  If the spins are half odd integer, the easy-plane configuration is not allowed, and then the ground state corresponds to the little ferromagnetic state where all the spins are aligned with $s=1/2$ or $s=-1/2$.  The anti-ferromagnetic case, $b>0$, is rather straightforward when $N$ is even and can be mapped directly to the ferromagnetic case using a symmetry between spins and staggered spins as given in Eqn.\eqref{st}.  For $N$ odd, the anti-ferromagnetic phase diagram is quite complicated, since the Nel state is always frustrated.  For $a>b\cos\frac{\pi}{N}$ the ground state is the easy-plane state with all spins $s=0$ for the case of integer spin, but for half odd integer spin where this state cannot be achieved, we get the little Nel states where the spins are $s=\pm 1/2$ but with one defect.  As the position of the defect is arbitrary, the ground state is then $N$ fold degenerate.  Then as we cross the line $a=b\cos\frac{\pi}{N}$ decreasing $a$ we encounter a ground state with a soliton.  The soliton is spread over the entire chain and interpolates between a  unit of the Nel state of length 2, with one spin that is maximally up, $\uparrow$, and its neighbour which is maximally down, $\downarrow$.  This state is of course $N$ fold degenerate as the position of the soliton is arbitrary.  There is also a two fold degeneracy as the two spin Nel configuration can be flipped.  Then as $a$ is further decreased towards zero, each time $a$ crosses $b\cos\frac{\pi}{m}$, $m=N,N-1,\dots,3$, the length of the soliton shortens by one lattice unit, and the length of the Nel part increases by one.  Finally, for $a<0$ we get the frustrated Nel state.  As the frustration can be anywhere, there is an $N$ fold degeneracy, and a two fold degeneracy as each configuration can be flipped.

The soliton exists as an excitation for the chain with an even number of sites and also for the open chain or the infinite chain.  The excitation energy of the soliton can be gapless, thus becoming the dominant excitation, which then would be more important than the usual spin wave excitations which are massive.  A detailed analysis of the properties of the solitons and the critical phenomena in this model will be presented in a forthcoming paper.

\section{Acknowledgments} 
We thank Ian Affleck and William Witczak-Krempa for useful discussions. M.P. thanks the Perimeter Institute for hospitality where this work was begun.   We thank  NSERC of Canada for financial support.   Research at Perimeter Institute is supported by the Government of Canada through Industry Canada and by the Province of Ontario through the Ministry of Research and  Innovation.



\begin{thebibliography}{99}

\bibitem{bc}M. Blume, Phys. Rev., 141:517Ð524, 1966; H. W. Capel, Physica, 32:966Ð988, 1966; {\it ibid}, 33:295Ð331, 1967; {\it ibid}, 37:423Ð441, 1967; Phys. Lett 23, 327,1966.
\bibitem{Ising}E. Ising, Z. Phys. 31: 253,  (1925).
 \bibitem{haldane} 
F. D. M. Haldane, \prl {\bf 50}, 1153 (1983)
\bibitem{bethe}
 H. Bethe, Z. Physik {\bf 71}, 205 (1931) 
  \bibitem{bethe2}
 L. Hulth\'en, Arkiv Mat. Astron. Fysik {\bf 26}A, 1 (1938)
 \bibitem{B}
 P. W. Anderson, Phys. Rev. {\bf 86}, 694 (1952)
 \bibitem{ml2001}
Meier F. and Daniel Loss, \prl {\bf 86}, 5373 (2001); Florian Meier, Jeremy Levy, Daniel Loss  \prb{\bf 68}, 134417 (2003)
\bibitem{simon}
Jonathan Simon,	 Waseem S. Bakr, Ruichao Ma, M. Eric Tai,	 Philipp M. Preiss	 and Markus Greiner, Nature {\bf 472}, 307 (2011) 
 \bibitem{vill}
 J. Villain, Physica {\bf 79}B, 1 (1975); F. Devreux and J. P. Boucher, J. Phys. Paris {\bf 48}, 1663 (1987); H.-J. Mikesha and M. Steiner, Adv. Phys. {\bf 40}, 191 (1991); 
Hans-Benjamin Braun and Daniel Loss, J. Appl. Phys. {\bf 79}, 6107 (1996); S. E. Nagler, W. J. L. Buyers, R. L. Armstrong, and B. Briat, \prl {\bf 49}, 590 (1982); N. Ishimura and H. Shiba, Prog. of Theo. Phys., {\bf 63}, 743 (1980)
\bibitem{bal}
L. Balents, Nature {\bf 464}, 199 (2010).
\bibitem{bin}
K. Binder and A. P. Young, \rmp {\bf 58}, 801
(1986).
\bibitem{kit}
A. Kitaev, Annals of Physics {\bf 321}, 2 (2006).
\bibitem{sm8}
 T. Holstein, and H. Primakoff, Phys. Rev. {\bf 58}, 1098 (1940); S. A Owerre, Can. J. Phys. {\bf 91}, 542 (2013) .
\bibitem{chud10}
E. M. Chudnovsky and Javier Tejada, Lectures on Magnetism with 128 problems.  Rinton Press, Princeton, NJ, (2006); E. M. Chudnovsky , Javier Tejada , Carlos Calero and Ferran Macia, Solutions to Lectures on Magnetism.  Rinton Press, Princeton, NJ, (2006); Gwang-Hee Kim, \prb {\bf 67}, 024421 (2003); ibid {\bf 68}, 144423 (2003)
\bibitem{klau}
John R. Klauder \prd{\bf 19}, 2349 (1978)
\bibitem{AB}
Alexander Altland and Ben Simons, Condensed Matter Field Theory, Cambridge University Press, New York, (2010); Hagen Kleinert, Path Integrals in Quantum
Mechanics, Statistics, Polymer Physics and Financial Markets, World scientific publishing Co. Pte. Ltd  (2009)
\bibitem{wesszu}
J. Wess, B. Zumino, Phys. Lett. B {\bf 37}:95,1971; S.P. Novikov,  Usp.Mat.Nauk,  37N5:3-49,1982; E. Witten, Nucl. Phys., B{\bf 160}:57,1979. 
\bibitem{sm2}
S. A Owerre  and M.B Paranjape, \prb {\bf 88}, 220403(R), (2013).
\bibitem{dav}
 Davis, Philip J., Circulant Matrices, Wiley, New York, 1970
 \bibitem{toeplitz}
  Devadutta Kulkarni, Darrell Schmidt and Sze-Kai Tsui, Lin. Alg. and Appl. (1999) 63-80; C.M. da Fonseca and J. Petronilho, Lin. Alg. and  Apps 325 (2001), 7-21.

\end{thebibliography}
 \end{document}